\documentclass[aps,english,preprintnumbers,amsmath,amssymb,superscriptaddress,twocolumn]{revtex4-2}
\pdfoutput=1

\usepackage{mathtools}
\usepackage{bbold}
\usepackage{amsfonts}
\usepackage{mathrsfs}
\usepackage{bbm}
\usepackage{slashed}

\usepackage{graphicx}
\usepackage[dvipsnames]{xcolor}
\usepackage{array}
\usepackage[percent]{overpic}
\usepackage{xspace}
\usepackage{hyperref}

\definecolor{blue}{RGB}{31, 119, 180}
\definecolor{orange}{RGB}{255, 127, 14}
\definecolor{green}{RGB}{44, 160, 44}
\definecolor{red}{RGB}{214, 39, 40}
\definecolor{purple}{RGB}{148, 103, 189}
\definecolor{brown}{RGB}{140, 86, 75}
\definecolor{pink}{RGB}{227, 119, 194}
\definecolor{gray}{RGB}{127, 127, 127}
\definecolor{olive}{RGB}{188, 189, 34}
\definecolor{cyan}{RGB}{23, 190, 207}

\usepackage{xifthen}
\usepackage{dsfont}
\usepackage{appendix}
\usepackage{enumitem}
\usepackage{booktabs}
\usepackage{units}
\usepackage[normalem]{ulem}
\usepackage{soul}

\makeatletter
\g@addto@macro\bfseries{\boldmath}
\makeatother

\setkeys{Gin}{width=0.48\textwidth}

\graphicspath{
	{./}
}

\def\0#1#2{\frac{#1}{#2}}

\def\CC{{\mathcal C}}

\DeclareMathOperator{\Tr}{Tr}

\newcommand{\be}{\begin{eqnarray}}
\newcommand{\ee}{\end{eqnarray}}

\newcommand{\beq}{\begin{equation}}
\newcommand{\eeq}{\end{equation}}
\newcommand{\bea}{\begin{eqnarray}}
\newcommand{\eea}{\end{eqnarray}}

\newcommand{\deltapq}[1]{\left(2 \pi\right)^4 \delta^{(4)}\left(#1\right)}

\newcommand{\mub}{\mu_\mathrm{B}}
\newcommand{\nb}{n_\mathrm{B}}
\newcommand{\ms}{m_\mathrm{s}}
\newcommand{\DeltaCFL}{\Delta_\mathrm{CFL}}
\newcommand{\barms}{\bar{m}_\mathrm{s}}
\newcommand{\barmue}{\bar{\mu}_\mathrm{e}}
\newcommand{\barmuthree}{\bar{\mu}_3}
\newcommand{\barmueight}{\bar{\mu}_8}
\newcommand{\barDeltaCFL}{\bar{\Delta}_\mathrm{CFL}}
\newcommand{\pfree}{p_\mathrm{free}}
\DeclareMathOperator{\diag}{diag}
\newcommand{\Seff}{S_\mathrm{eff}}
\newcommand{\muL}{\mu_{\rm L}}
\newcommand{\nL}{n_{\rm L}}
\newcommand{\pL}{p_{\rm L}}
\newcommand{\muH}{\mu_{\rm H}}
\newcommand{\nH}{n_{\rm H}}
\newcommand{\pH}{p_{\rm H}}
\newcommand{\cs}{c_\mathrm{s}}
\newcommand{\csSqext}{c_\text{s,ext}^2}

\def\0#1#2{\frac{#1}{#2}}

\newcommand{\orcid}[1]{\href{https://orcid.org/#1}{\includegraphics[height=1.9ex,width=1.9ex]{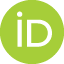}}}

\hypersetup{
	colorlinks,
	linkcolor={red!75!black},
	citecolor={blue!75!black},
	urlcolor={blue!75!black},
	pdftitle={Color superconductivity under neutron-star conditions at next-to-leading order},
	pdfauthor={Geißel, Gorda, Braun},
	bookmarksopen=true,
	bookmarksopenlevel=2,
	bookmarksnumbered=true
}

\usepackage{babel}
\makeatother
\begin{document}

\title{\texorpdfstring{Color superconductivity under neutron-star conditions at next-to-leading order}{Color superconductivity under neutron-star conditions at next-to-leading order}}

\author{Andreas Gei\ss el \orcid{0009-0007-9283-4211}\,}
\email{andreas.geissel@tu-darmstadt.de}
\affiliation{Institut f\"ur Kernphysik, Technische Universit\"at Darmstadt,
	64289 Darmstadt, Germany}
\author{Tyler Gorda \orcid{0000-0003-3469-7574}\,}
\email{gorda@itp.uni-frankfurt.de}
\affiliation{Institut f\"ur Theoretische Physik, Goethe Universit\"at,   Max-von-Laue-Straße 1, 60438 Frankfurt am Main, Germany}
\author{Jens Braun \orcid{0000-0003-4655-9072}\,}
\email{jens.braun@physik.tu-darmstadt.de}
\affiliation{Institut f\"ur Kernphysik, Technische Universit\"at Darmstadt, 64289 Darmstadt, Germany}
\affiliation{ExtreMe Matter Institute EMMI, GSI Helmholtzzentrum f\"ur Schwerionenforschung GmbH, Planckstraße 1, 64291 Darmstadt, Germany}
\begin{abstract}
	The equation of state of deconfined strongly interacting matter at high densities remains an open question, with effects from quark pairing in the preferred color-flavor-locked (CFL) ground state possibly playing an important role. 
	Recent studies suggest that at least large pairing gaps in the CFL phase are incompatible with current astrophysical observations of neutron stars. 
 	At the same time, it has recently been shown that in two-flavor quark matter, subleading corrections from pairing effects can be much larger than would be na\"ively expected, even for comparatively small gaps. 
	In the present~\textit{Letter}, we compute next-to-leading-order corrections to the pressure of quark matter in the CFL phase arising from the gap and the strong coupling constant, incorporating neutron-star equilibrium conditions and current state-of-the-art perturbative QCD results.
	We find that the corrections are again quite sizable, and they allow us to constrain the CFL gap in the quark energy spectrum to $\Delta_{\rm CFL} \lesssim 140$~MeV at a baryon chemical potential $\mub = 2.6~$GeV, even when allowing for a wide range of possible behaviors for the dependence of the gap on the chemical potential. 
\end{abstract}
\maketitle
{\it Introduction.--}
The equation of state (EOS) of strong-interaction matter at low temperatures and high densities remains an open question in high-energy and nuclear physics.
First-principles lattice calculations of quantum chromodynamics~(QCD) fail in this regime due to the sign problem~\cite{deForcrand:2009zkb, Philipsen:2012nu, Aarts:2015tyj, Gattringer:2016kco, Nagata:2021ugx}.
Nevertheless, recent observations of binary neutron-star (NS) merger events~\cite{TheLIGOScientific:2017qsa, LIGOScientific:2018cki, LIGOScientific:2018hze, LIGOScientific:2020aai}, as well as other astrophysical measurements of NSs~\cite{Antoniadis:2013pzd, Cromartie:2019kug, Fonseca:2021wxt, Steiner:2017vmg, Nattila:2017wtj, Shawn:2018, Miller:2019cac, Riley:2019yda, Miller:2021qha, Riley:2021pdl, Choudhury:2024xbk, Saffer:2024tlb},
have revitalized interest in the thermodynamics of dense QCD matter at low temperatures.

In particular, there has been a renewed focus on the deconfined phase of QCD in recent years.
This has been driven by a push to complete the full next-to-next-to-next-to-leading-order computation of the EOS using effective-field-theory~(EFT) techniques~\cite{Gorda:2018gpy, Gorda:2021kme, Gorda:2021znl, Gorda:2023mkk, Karkkainen:2025nkz} and renewed interest in the role of these perturbative results in the context of NS EOS inference~\cite{Komoltsev:2021jzg, Gorda:2022jvk, Somasundaram:2022ztm, Komoltsev:2023zor, Zhou:2023zrm} and the application of lattice QCD results in QCD-like theories~\cite{Brandt:2022hwy, Abbott:2023coj, Abbott:2024vhj} to NS EOS inference~\cite{Moore:2023glb, Fujimoto:2023unl, Navarrete:2024zgz}.
A key unknown that has recently been increasingly investigated in dense QCD matter is the role of the non-perturbative quark pairing in the ground state~\cite{Leonhardt:2019fua, Braun:2021uua, Braun:2022jme, Fujimoto:2023mvc, Geissel:2024nmx, Kurkela:2024xfh, Fukushima:2024gmp}.

The condensation of diquarks via attractive gluon exchange in cold quark matter has long been expected~\cite{Barrois:1977xd, Bailin:1983bm, Alford:1997zt, Son:1998uk, Rapp:1997zu, Schafer:1999jg, Pisarski:1999tv}.
At large baryon chemical potential~$\mu_{\rm B}$, where the three lightest quark flavors are active and flavor breaking is suppressed, the dominant pairing channel is color-flavor locking (CFL). 
The corresponding diquark condensate takes the form
\begin{align}
\langle\psi_a^i \mathcal{C}\gamma_5 \psi_b^j \rangle\sim\DeltaCFL \epsilon_{ab A}\epsilon^{ij A} \, ,
\end{align}
where~$a,b$ are color and~$i,j$ flavor indices ~\cite{Alford:1997zt,Rapp:1997zu,Alford:1998mk,Alford:2002kj}.
The summation over the index~$A$ ``locks" color and flavor indices in a specific pattern and implies that the CFL condensate breaks chiral symmetry.
General considerations and model calculations indicate that this channel dominates over the less symmetric two-color-superconducting (2SC) channel, in which only two quark flavors participate, see~\cite{Rajagopal:2000wf, Rischke:2003mt, Buballa:2003qv, Alford:2007xm} for reviews and~\cite{Gholami:2024ety} for recent developments.
At leading order~(LO) in the strong coupling and~$\mathcal{O}(\Delta_{\rm CFL}^2)$, the CFL condensate shifts the pressure above the unpaired pressure of normal quark matter~(NQM) by the condensation energy 
\begin{equation}
\label{eq:pressurepairedLO}
p_\mathrm{paired} = \frac{\mub^2\DeltaCFL^2}{3 \pi^2}\,,
\end{equation}
where the gap~$\DeltaCFL$ carries a non-analytic dependence on the strong coupling~$\alpha_s$.

As recently noted by Kurkela, Steinhorst, and Rajagopal, assuming a constant gap, the LO pressure shift can already lead to tension with current astrophysical observations if the gap is too large~\cite{Kurkela:2024xfh}.
However, NLO corrections in~$\alpha_s$ may be sizable~\cite{Geissel:2024nmx}, and the gap's scaling with~$\mu_{\rm B}$ significantly affects derived quantities like the speed of sound.
Incorporating these corrections is thus essential to obtain realistic gap constraints.

In this \emph{Letter}, we present NLO results for the pressure in the CFL phase in both the strong coupling and the strange quark mass~$\ms$, the latter of which must be included because it competes with the gap.
To make contact with applications within the context of NS EOS inference, we study a system in equilibrium under the strong, weak, and electromagnetic forces (dubbed ``NS conditions'' below).
Combining these results with the current state-of-the-art NQM perturbative-QCD (pQCD) ones with nonzero strange quark mass~\cite{Kurkela:2009gj, Gorda:2021gha}, we obtain a precise, first-principles description of cold and dense quark matter that allows us to constrain the size of the pairing gap even when allowing for a wide range of possible behaviors for the dependence of the gap on the chemical potential.

{\it Pressure of CFL quark matter.--}
We employ our framework from Ref.~\cite{Geissel:2024nmx} to systematically compute the coefficients of an expansion of the pressure~$p$ at zero temperature and large chemical potentials. 
Defining~${\barms \equiv \ms/(\mub/3)}$,~${\barDeltaCFL \equiv \DeltaCFL/(\mub/3)}$ for the small quantities in our approximation, we may write the pressure in the form
\begin{align}
\label{eq:pressureExpansion}
p = \pfree \bigl[\gamma_0(\alpha_s, \barms^2) + \gamma_1(\alpha_s, \barms^2)\barDeltaCFL^2 + \dots\bigr]\,,
\end{align}
where~$\pfree \equiv \mub^4 / (108 \pi^2)$ is the free pressure of three massless quark flavors with equal chemical potentials. 
The~$\gamma_0$ term gives the NQM result; we aim to compute corrections to~$\gamma_1$ in the small quantities~$\alpha_s$ and~$\barms^2$.
Note that~$\barms \ll 1$ is a valid approximation at high densities where QCD is weakly coupled~\cite{Gorda:2021gha}.

Let us first discuss the chemical potentials involved.
The CFL condensate involves all nine quark flavors and colors in a particular pattern.
Six quarks have a unique pairing partner and the remaining three pair mutually.
Since the strange-quark mass breaks flavor symmetry explicitly, we introduce different chemical potentials for each color and flavor.
Following Ref.~\cite{Alford:2002kj}, for CFL matter that is both charge and color neutral, the chemical-potential matrix reads
\begin{align}
\label{eq:mufc}
\hat\mu_{\rm f,c} =  \mub/3 -  \mu_{\rm e} Q_{\rm f,c} + \mu_3 T^3_{\rm f,c} + \mu_8 T^8_{\rm f,c}\,,
\end{align}
with flavor and color indices~$\rm f$ and~$\rm c$.
The first term arises from baryon number conservation,~${Q=\diag(2/3,-1/3,-1/3)} \otimes 1_\mathrm{c}$ is the electric charge matrix, and~${T^3=1_\mathrm{f} \otimes \diag(1/2,-1/2,0)}$ and~${T^8= 1_\mathrm{f} \otimes \diag(1/3,1/3,-2/3)}$ correspond to the generators of the SU(3) gauge group that characterize possible color-neutral pairings of the quarks.
The latter three conserved quantities have associated chemical potentials~$\mu_\mathrm{e}$,~$\mu_3$, and~$\mu_8$ respectively.
The quarks that are allowed to pair form common Fermi momenta to minimize the free energy, if the gap is not too small compared to the strange quark mass.
At LO, the gap must satisfy~$\barDeltaCFL > \barms^2/4$~\cite{Alford:2002kj}.
Hence, for a finite strange quark mass~$\ms$, there are four different common Fermi momenta, three for the uniquely paired quarks and one for the remaining three quarks pairing mutually.

The chemical-potential matrix makes loop calculations quite complicated.
However, under NS conditions, an important simplification arises.
Since for vanishing strange quark mass~$\ms \to 0$, three-flavor quark matter with equal chemical potentials already satisfies the NS conditions, we see that for small~$\barms$,~$\barmue \equiv \mu_{\rm e}/(\mub/3)$,~$\barmuthree \equiv \mu_{3}/(\mub/3)$, and~$\barmueight \equiv \mu_{8}/(\mub/3)$ are all parametrically small.
As shown in Ref.~\cite{Alford:2002kj},~$\barmue \sim \barmuthree \sim \barmueight \sim \barms^2$.
Thus, we can consistently expand also in these additional chemical potentials.
We show explicitly in App.~\ref{app:colorneutralityimplieschargeneutrality} that in color-neutral CFL matter~$\bar\mu_{\rm e} = 0$, which follows from a simple argument about the number densities of quarks~\cite{MichaelBuballa, Rajagopal:2000ff}. 

Our approach to compute corrections to the coefficient~$\gamma_1$ begins with the effective action for QCD in the CFL phase
\begin{align}
\label{eq:Seff}
\Seff \equiv \int_x \Big\{& \bar{\psi} \left({\rm i}\slashed{D} - {\rm i} \hat\mu_{\rm f,c} \gamma_0 - \hat M_{\rm f,c} \right)\psi + \frac{1}{4} F_{\mu\nu}^a F_{\mu\nu}^a \\
&+ m^2 \Delta^{2} \nonumber+ \frac{1}{2}(\psi^T\CC\gamma_5\Delta \epsilon_{a}^{\rm f}\epsilon_{a}^{\rm c}\psi)  + \text{h.c.} \Big\} \,.
\end{align}
Here, color and flavor indices on the quark fields~$\psi$ are suppressed,~$\Delta$ represents the diquark field with a mass parameter~$m$,~${\hat M_{\rm f,c} \equiv \diag(0,0,\ms) \otimes 1_{\rm c}}$ is the quark mass matrix, and~$\epsilon_{a}^{\rm f}$ and~$\epsilon_{a}^{\rm c}$ are totally antisymmetric matrices in flavor and color space, respectively, with~$a\in \lbrace1,2,3\rbrace$ locking color and flavor indices.
The effective action~\eqref{eq:Seff} is motivated by the same logic as in Ref.~\cite{Geissel:2024nmx}.
It follows from considering an infinitesimal renormalization-group step from the bare QCD Lagrangian, followed by an (exact) Hubbard-Stratonovich transformation trading the four-fermion interaction for a diquark-fermion-fermion interaction.
An infinite number of other terms are induced by this, but only those in Eq.~\eqref{eq:Seff} are necessary to the order we are working.
For our computation,~$\Delta$ is taken as a constant background field.
Its explicit dependence on~$\mub$ will be fixed by minimizing the effective action and solving for a particular mass~$m$ for the diquark field.

The propagators from~$\Seff$ involve all combinations of~$\psi$,~$\psi^T$ and their Dirac adjoints and are furthermore nondiagonal in color and flavor space due to the chemical potential matrix~$\hat \mu_{\rm f,c}$ in Eq.~\eqref{eq:mufc}.
To compute~$\gamma_1$ to the order of interest, we must compute the quantum effective action~$\Gamma$, which is the Legendre transform of the logarithm of the partition function, to two-loop level.
The pressure in Eq.~\eqref{eq:pressureExpansion} then follows from an evaluation of~$\Gamma$ at its ground state~$\Delta=\DeltaCFL$, remembering that we can expand in the small quantities~$\barms$ and~$\barDeltaCFL$.

Details of the computation at one-loop order with a finite strange-quark mass are given in App.~\ref{app:strangequarkmass}. 
For the two-loop corrections, there is no mixing between the~$\barms$ and~$\barDeltaCFL$ since such terms are higher order. 
Hence the corrections can be computed separately.
Mass-dependent corrections follow the expansion scheme of Ref.~\cite{Gorda:2021gha}, including now the chemical-potential matrix~$\hat\mu_{\rm f, c}$.
For the gap corrections, each propagator is split as
\begin{equation}
\mathcal{P}_\psi =  \mathcal{P}_\psi^0 + (\mathcal{P}_\psi - \mathcal{P}_\psi^0) \equiv \mathcal{P}_\psi^0 + \mathcal{P}_\psi^\Delta\,,
\end{equation}
where~$\mathcal{P}_\psi^0 =\mathcal{P}_\psi|_{\Delta = 0}$, and we consider diagrams expanded up to one term in the gapped propagators~$\mathcal P _\psi^\Delta$, as in~\cite{Geissel:2024nmx},
\begin{align}
\label{eq:2loopdiagamsapprox}
\hspace{.212cm}
\begin{aligned}
\includegraphics[width=0.4\textwidth]{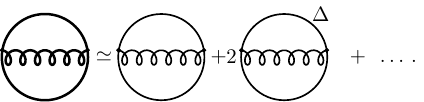}
\end{aligned}
\end{align}
Thick lines denote full propagators, while thin lines correspond to~$\mathcal{P}_\psi^0$ and thin lines labeled with~$\Delta$ are associated with gapped corrections.
Since~$\barmue$,~$\barmuthree$, and~$\barmueight$ are small, this diagram can be computed with techniques from Ref.~\cite{Geissel:2024nmx}; we provide some details in Apps.~\ref{app:FeynmanRules} and~\ref{app:twoloopcontribution}.

After computing the diagrams shown in Eq.~\eqref{eq:2loopdiagamsapprox}, we find up to second order in~$\bar\Delta_{\rm CFL}$,
\begin{align}
\label{eq:pressureNLO}
\frac{p}{p_{\rm free}} ={}& 1 + \frac{2}{9} (\barmue^2 +\barmueight^2 - \barmuthree\barmueight) +\frac{1}{6}(\barmuthree^2 - 2\barmuthree\barmue)
\nonumber\\
&- \barms^2 + \frac{\barms^2}{9}(2 \barmueight - \barmue) + \frac{7\barms^4}{72} - \frac{\barms^4}{2}\ln\Bigl(\frac{\barms}{2}\Bigr)
\nonumber\\
&- \frac{2 \alpha_s}{\pi} - \frac{2 \alpha_s \barms^2}{3 \pi} \biggl\{ 5 + 6 \ln \biggl[\frac{\Lambda}{2(\mub / 3)}\biggr] \biggr\}
\nonumber\\
&+ 4 \barDeltaCFL^2 - \frac{4}{3}\barms^2 \barDeltaCFL^2 +  40.9 \alpha_s \barDeltaCFL^2\,.
\end{align}
Here,~$\Lambda$ denotes the renormalization scale in the modified minimal subtraction scheme.
Note that, in principle, two terms are missing: one proportional to~$\alpha_s \barmueight$ and another~$\barmueight \barDeltaCFL^2$, which, however, do not appreciably impact our results (see below).
From the gap-dependent terms we extract the expansion coefficient~$\gamma_1$ in Eq.~\eqref{eq:pressureExpansion},
\begin{equation}
{\gamma_1(\alpha_s, \barms^2) = 4 - 4\barms^2/3 +  40.9\alpha_s\, .}
\end{equation}
This is one of our main results.
We would like to add that Goldstone modes associated with the symmetry breaking pattern of the CFL phase~\cite{Alford:2007xm} do not contribute to~$\gamma_1$ at~$\mathcal O (\alpha_s)$ but only effectively enter at higher orders, e.g., via two-gluon exchange subgraphs between quarks.

Specializing to NS conditions, we add the pressure of a non-interacting electron gas~$p_{\rm e} = \mu_{\rm e}^4 /(12\pi^2)$ and impose neutrality under the electromagnetic and strong forces, viz.~${\partial p / \partial \mu_{\rm e} = \partial p / \partial \mu_{\rm 3} = \partial p / \partial \mu_{\rm 8} = 0}$, to fix the relevant chemical potentials.
These conditions imply~$\barmue = \barmuthree = 0$ and~$\quad \barmueight = -\barms^2/2$~\cite{Alford:2002kj}, showing that the remaining chemical potentials are parametrically small.
Substituting this into Eq.~\eqref{eq:pressureNLO}, we obtain the pressure of quark matter under NS conditions
\begin{align}
\label{eq:pressureNLOneutral}
p^{\rm NS}_{\rm CFL} &= p^{\rm NS}_{\rm NQM} + p_{\rm free} \biggl[ 
\gamma_1(\alpha_s,m_{\rm s}^2)
\barDeltaCFL^2 - \frac{\barms^4}{4} \biggr] \, ,
\end{align}
with~$p^{\rm NS}_{\rm NQM}$ the pressure of NQM under NS conditions to two-loop level~\cite{Fraga:2004gz}
\begin{align}
p^{\rm NS}_{\rm NQM} ={}& p_{\rm free} \biggl( 1 - \bar m_{\rm s}^2 +\frac{7-12\ln\left(\bar m_{\rm s}/2\right)}{24}\bar m_{\rm s}^4 \nonumber\\
&\!\!\!\!\!\!- \frac{2 \alpha_s}{\pi} - \frac{2 \alpha_s \barms^2}{3 \pi} \biggl\{ 5 + 6 \ln \biggl[\frac{\Lambda}{2 (\mub / 3)}\biggr] \biggr\} \biggr)
\end{align}
Here, the constant independent of~$\barms$ and~$\alpha_{s}$ reproduces the condensation energy in Eq.~\eqref{eq:pressurepairedLO}.
From this, we directly deduce a criterion for whether the CFL phase is favored over NQM.
Comparing the pressure of NQM in NS conditions with the pressure of CFL matter in Eq.~\eqref{eq:pressureNLOneutral}, we see the CFL phase is favored if
\begin{align}
\label{eq:CFLfavoredconditionNLO}
\barDeltaCFL >  \frac{\barms^2}{4}\bigg[ 1 - 5.11\alpha_s + \frac{\barms^2}{6}\bigg] \, ,
\end{align}
where the sum of the subleading corrections is negative where the result is converged.
Hence, Eqs.~\eqref{eq:pressureNLOneutral} and \eqref{eq:CFLfavoredconditionNLO} show that the NLO corrections further increase the pressure in the CFL phase, and lower the minimal gap required for stability at nonvanishing~$\barms$.

From the EOS, we can also gain insight into the competition between 2SC and CFL at high densities.  
Indeed, by computing the coefficient~$\gamma_1$ for the 2SC phase with three massless flavors along the lines of Ref.~\cite{Geissel:2024nmx} and comparing 
the result,~${\gamma_1^{\text{2SC}} \left(\alpha_{s}\right) = 4/3 + 9.17\alpha_{s}}$, to our main result for CFL matter above, we conclude that the NLO corrections render the CFL ground state even more stable than expected from a mean-field analysis~\cite{Alford:2002kj}. 
In other words, for the 2SC ground state to be realized, the corresponding gap must be significantly greater than the CFL gap to compensate~$\gamma_1^{\text{2SC}}$. 
To be specific, for~$m_{\rm s}=0$, we find the 2SC phase to be dominant if~${\Delta_{\text{2SC}}/\Delta_{\text{CFL}}>\sqrt{3} + 2.90\alpha_{s}}$. 
This may suggest that CFL-type pairing remains favored even down to the nucleonic regime.

{\it Constraining the CFL gap and applications.--}
We now use our expression for~$p^{\rm NS}_{\rm CFL}$ to place updated bounds on the color-superconducting gap.
As our corrections shift the pressure to higher values, the new bounds will be tighter than those of Ref.~\cite{Kurkela:2024xfh}.
For this, we replace~$p_{\rm NQM}^{\rm NS}$ derived above with the state-of-the-art pQCD pressure at order~${\mathcal O}(\alpha_s^{5/2})$~\cite{Gorda:2021gha}, assuming small coupling and quark mass~\footnote{This work uses the fact that~$\barms^2$ and~$\alpha_s$ are terms of similar size where the pQCD results are reliable.}.
Explicitly, we add to our above results a term of~${\mathcal O}(\alpha_s^2)$ independent of~$\bar m_{\rm s}^2$~\cite{Freedman:1976xs, Freedman:1976ub}, which effectively includes LO Goldstone effects.
We further assume a power-law scaling of the gap,
\begin{align}
\label{eq:gap_scaling}
\DeltaCFL  = \DeltaCFL^{*} \left(\frac{\mub}{\mub^*}\right)^\sigma \,,
\end{align}
with constant exponent~$\sigma$, and gap size~$\DeltaCFL^{*}$ at reference chemical potential~$\mub^*$. 
This has been shown to be a good approximation for the gap as derived using both weak-coupling techniques and the functional renormalization group (fRG)~\cite{Braun:2022jme, Geissel:2024nmx} -- though the constants differ.
Typical values are~$\sigma \approx -0.23$ in the weak-coupling case~\cite{Son:1998uk} and~$\sigma \approx 0.45$ from fRG~\cite{frgCFLforthcoming}. 

An upper bound on the gap at high densities given information about the EOS at low densities can be understood as follows~\cite{Komoltsev:2021jzg, Gorda:2022jvk, Kurkela:2024xfh}:
Consider an EOS passing through two points~$(\muL, \nL, \pL)$ and~$(\muH, \nH, \pH)$ with chemical potential~$\muL < \muH$.
Since the speed of sound is related to the baryon density~$\nb$ by~${\cs^2 = (\nb/\mu)/ (\partial \nb / \partial \mub)}$ and satisfies~$0 \leq \cs^2 \leq 1$, one finds a maximal pressure difference along the EOS from~$\muL$ to~$\muH$:
\begin{equation}
\pH - \pL \leq \frac{\nH}{2}\left(\muH-\frac{\muL^2}{\muH}\right)\,.
\end{equation}
As the NS-matter EOS is known at low chemical potentials, e.g.,~from chiral EFT~\cite{Tews:2012fj, Hebeler:2013nza, Lynn:2015jua, Drischler:2017wtt, Drischler:2020hwi, Keller:2022crb} and astrophysical data, one can place an upper bound on the pressure and hence the gap at large chemical potentials.

Using these considerations, we derive an analytic bound on the gap at LO:
\begin{equation}
\label{eq:analyticestimateGapconstrain}
\Delta_{\rm CFL}^2(\mu_{\rm H}) \leq \frac{3\pi^2}{2}\frac{\muH n_{\rm NQM}^{\rm NS}(\mu_{\rm H}) - 2 p_{\rm NQM}^{\rm NS}(\muH)}{\mu_L^2-\sigma \mu_{\rm H}^2}\,,
\end{equation}
assuming~$\pL \ll p_{\rm NQM}^{\rm NS}(\muH)$ and~$\muL^2 \ll \muH^2$ to simplify the expression.
We hence see that the~$\mub$-dependence of the gap significantly affects the constraint.
The constraint becomes weaker (stronger) for positive (negative) values of~$\sigma$.
In particular, for~$\sigma \geq \muL^2/\muH^2$, no bound is obtained at all.
Note that our result in Eq.~\eqref{eq:analyticestimateGapconstrain} represents a generalization of the bound found in Ref.~\cite{Kurkela:2024xfh}, where the gap was assumed constant.

\begin{figure}[t]
	\hspace{-1cm}\includegraphics[width=0.65\linewidth]{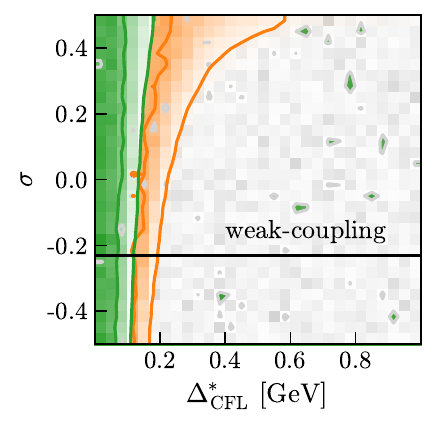}
	\caption{Two-dimensional prior~(gray) and posterior distributions~(orange, green) for the magnitude of the CFL gap~$\DeltaCFL^*$ and the scaling parameter~$\sigma$.
		The orange posterior corresponds to the conservative ensemble, while the green corresponds to the symmetric one (see main text).}
	\label{fig:PriorPosteriorDistribution}
\end{figure}

We turn to a Bayesian determination of~$\DeltaCFL^{*}$ values consistent with current astrophysical observations and our NLO EOS.
For the pQCD input, we take a log-uniform prior on the renormalization scale~$\Lambda/(2\mub/3) \in [1/2, 2]$ appearing in the NQM pressure, and fix~$\mub^* = 2.6$~GeV as the reference scale.
We also draw~$\sigma \in [-0.5, 0.5]$ uniformly, spanning both the weak-coupling and fRG values and take a uniform prior on~${\DeltaCFL^{*} \in [0,1]}$~GeV.
For each draw of the high-density pQCD information, we marginalize over chiral EFT and astrophysical information at lower densities by marginalizing over the astrophysical posterior from Ref.~\cite{Gorda:2022jvk}.
This posterior incorporates chiral-EFT below~$1.1$ nuclear saturation density $n_0\approx 0.16~\text{baryons}/\text{fm}^{3}$~\cite{Hebeler:2013nza}, and various astrophysical information.
We use mass measurements of PSR J0348$+$0432~\cite{Antoniadis:2013pzd} and PSR J1624$-$2230~\cite{Fonseca:2021wxt}; simultaneous mass and radius posteriors of PSR J0740$+$6620 from the NICER collaboration~\cite{Miller:2021qha}, the tidal deformability information from GW170817~\cite{LIGOScientific:2018hze}; and the assumption that a black hole was formed in GW170817, due to the observation of an electromagnetic counterpart \cite{LIGOScientific:2017ync, Margalit:2017dij, Shibata:2017xdx, Rezzolla:2017aly, Ruiz:2017due, Shibata:2019ctb}.
Within this marginalization, we additionally exclude combinations of low- and high-density EOSs that cannot be connected by a causal, stable, and thermodynamically consistent EOS extension without exceeding a given maximum speed of sound squared~$\csSqext$~\cite{Gorda:2022jvk}.

This marginalization involves two free parameters: the low-density matching point~$\nL$, and~$\csSqext$.
In Ref.~\cite{Kurkela:2024xfh}, where such an analysis has been performed without~$\mathcal O (\alpha_{s})$ corrections to~$\gamma_1$ and the chemical-potential dependence of the gap, a maximally ``conservative'' ensemble was defined: taking~$\nL = n_{2.1 M_\odot}$, the density reached in a 2.1-solar-mass NS, and~$\csSqext = 1$.
Here, we consider this ensemble as well as a ``symmetric'' ensemble.
For the latter we take~$\nL = n_\mathrm{TOV}$, the maximum density reached in a stable NS, and~${\csSqext = 2/3}$.
We take~$\csSqext$ so that~$\cs^2$ in the interpolated region can be above or below the conformal value of~$1/3$ with equal probability for a uniform prior.
In Fig.\ref{fig:PriorPosteriorDistribution}, these are shown with 68\% and 95\% credibility contours: orange for conservative, green for symmetric.
We use the running values of $\alpha_s$ and $\ms$ at next-to-next-to-next-to-leading order, and fix the scales by setting~${\alpha_s(2\,\text{GeV}) = 0.2994}$~\cite{ParticleDataGroup:2008zun} and~${\ms(2\,\text{GeV}) = 93.8\,\text{MeV}}$~\cite{Choudhury:2024xbk}.
As predicted analytically, the constraint on~$\DeltaCFL^{*}$ becomes weaker with increasing~$\sigma$, though less so for the symmetric ensemble than the conservative one.
Compared to the LO results, NLO corrections tighten the 95\% upper bound by about a factor of two.
After marginalizing over all the other parameters, we find~$\DeltaCFL^{*} \lesssim 140$~MeV at 95\% credibility for the symmetric ensemble.

To test the robustness of our results against omitted terms in Eq.\eqref{eq:pressureNLO}, we varied the coefficients of the~$\mathcal{O}(\alpha_s \barms^2)$ and~$\mathcal{O}(\barms^2 \barDeltaCFL^2)$ terms by a factor of two, accounting for both the neglected terms and the mixed renormalization scheme.
This variation induced a shift of less than 0.5\% in the upper gap bound, confirming their negligible impact.

\begin{figure*}[t]
	\centering
	\includegraphics[width=0.45\textwidth]{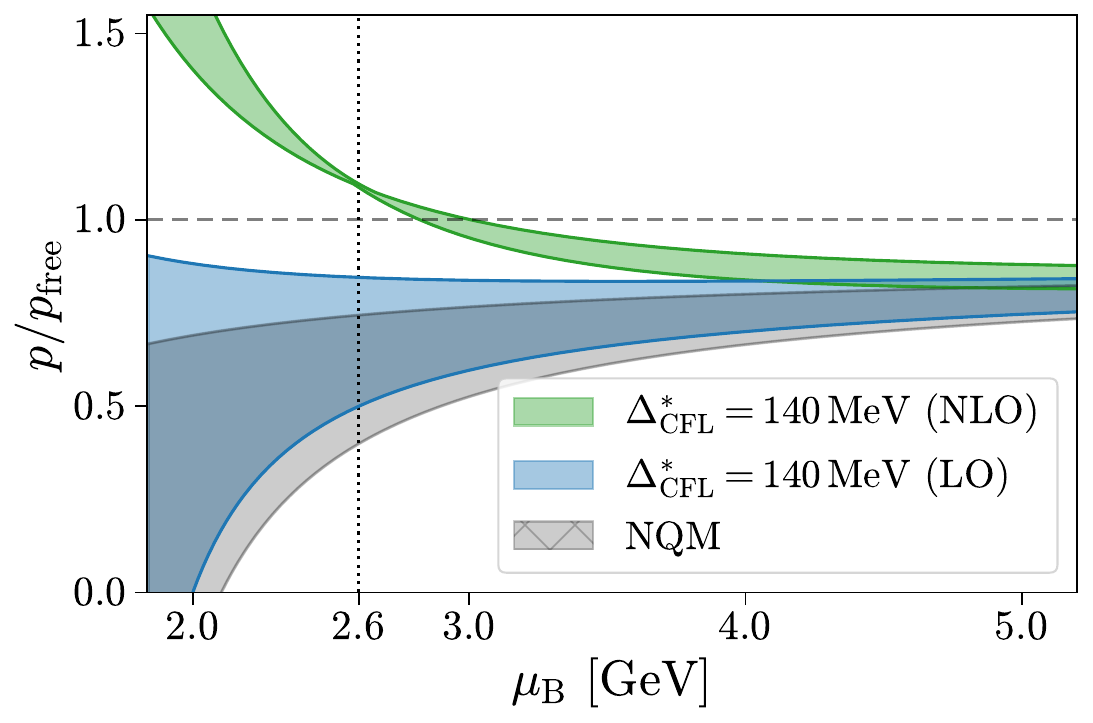}
	\hfil
	\includegraphics[width=0.45\textwidth]{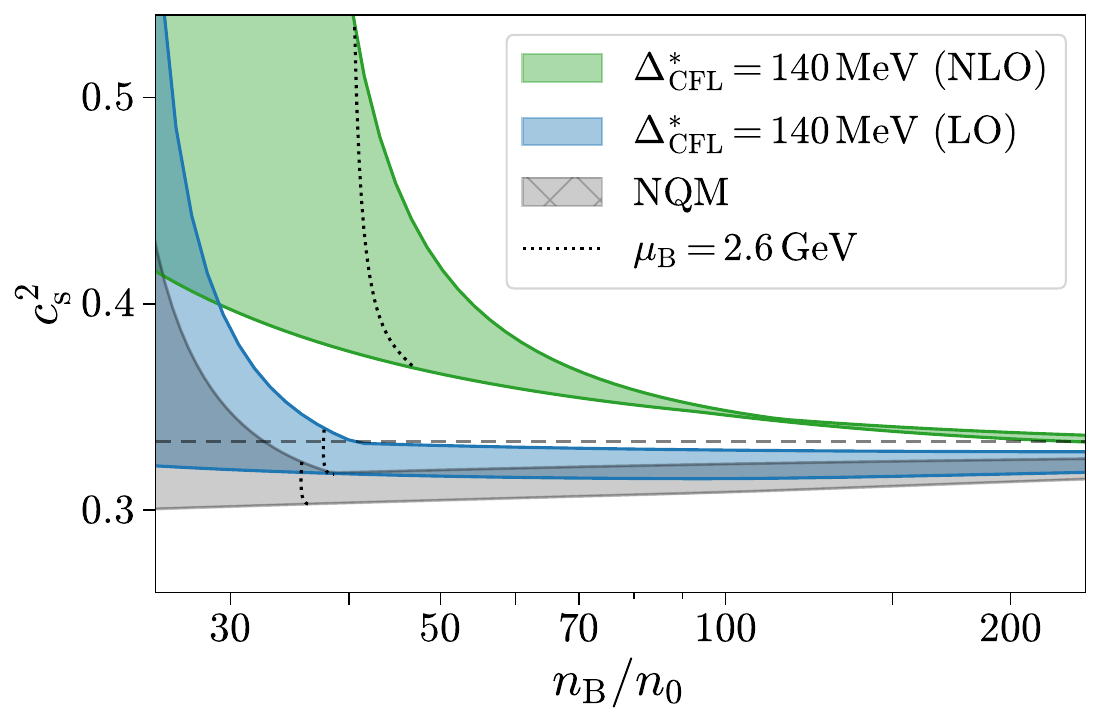}
	\caption{Left: Normalized pressure as a function of baryon chemical potential. Right: Speed of sound as a function of baryon density in units of nuclear saturation density. 
		The dotted lines in both panels denote the baryon chemical potential at~$\mub^*= 2.6$~GeV.
		The uncertainty bands stem from the usual renormalization-scale variation (see main text).}
	\label{fig:p_cs2}
\end{figure*}

Selecting this 95\% credible bound on the gap, in Fig.~\ref{fig:p_cs2} we show the normalized pressure as a function of~$\mub$ and the speed of sound as a function of the baryon density~$\nb$, for the weak-coupling scaling~$\sigma = -0.23$.
The NLO corrections significantly enhance both quantities beyond the LO correction from the gap.
The normalized pressure and~$\cs^2$ both still approach their free values at large~$\mub$ and~$\nb$ from below but can differ sizably in the NLO case, even at densities where the renormalization-scale variation errors are small.
Notably, the NLO speed of sound begins to exceed the conformal value even above~$\nb = 200 n_0$, and is only well converged down to about~$50 n_0$, in contrast to the NQM results~\cite{Komoltsev:2023zor}.
Though our figures are for a very large value of the gap consistent with current astrophysical data, these results suggest that the weak-coupling expansion is under poor control once corrections from the gap and coupling are taken into account.

{\it Conclusion and outlook.--}
In this \emph{Letter}, we have presented an NLO expansion of the pressure of dense CFL matter at zero temperature.
In particular, we have calculated NLO corrections in the CFL gap~$\DeltaCFL$, the strong coupling~$\alpha_s$, and the strange quark mass~$\ms^2$ both in the general case, and under NS conditions, i.e., equilibrium under the strong, weak, and electromagnetic forces.
We have shown that these NLO corrections provide further stability to CFL pairing as compared to 2SC pairing of three-flavor quark matter at high densities.
By folding in information from current astrophysical observations of NSs and their mergers, we have placed an upper bound on the gap of~${\DeltaCFL(\mub = 2.6\,\text{GeV}) \lesssim 140\,\text{MeV}}$ at pQCD densities using our new results and while allowing for a wide range of possible behaviors of the superconducting gap as a function of~$\mub$.
This constitutes a strong constraint on non-perturbative and model-based estimates of the gap.
Moreover, since our NLO corrections are very sizable, our results suggest that the weak-coupling expansion of the pressure may converge more poorly at high densities in the physical pairing channel than previously suggested for the normal phase.
Clearly, further calculations in the paired phase of high-density quark matter are necessary to resolve this apparent discrepancy and provide a converged EOS over a wide density range.

{\it Acknowledgments.--~}
We thank Michael Buballa, Aleksi Kurkela, and Sanjay Reddy for helpful discussions.
This work has been supported in part by the Deutsche Forschungsgemeinschaft (DFG, German Research Foundation) project-ID 279384907 -- SFB 1245 (J.B., A.G.), 
by the DFG project 315477589 -- SFB TRR 211 (J.B., T.G.),
by the State of Hesse within the Research Cluster ELEMENTS (projectID 500/10.006) (J.B., T.G.), 
and by the ERC Advanced Grant ``JETSET: Launching, propagation and emission of relativistic jets from binary mergers and across mass scales'' (Grant No.~884631) (T.G.).

{\it Data availability.--}
The data that support the findings of this article are openly available~\cite{Geissel:2025vnpData}.

\bibliography{refs}


\appendix

\begin{widetext}
\section*{End Matter}
\end{widetext}

\section{Color neutrality implies charge neutrality}
\label{app:colorneutralityimplieschargeneutrality}
In this appendix, we demonstrate that color-neutral CFL matter is intrinsically electrically neutral, even in the absence of electrons, i.e.,~$\bar\mu_{\rm e} = 0$~\cite{Buballa:2003qv, Rajagopal:2000ff}.
Due to the specific pairing patterns among quarks of different colors and flavors in CFL matter, certain density relations hold~\cite{MichaelBuballa}:
$
n_{rd} =  \,\,n_{gu}\,, n_{rs} = n_{bu}\,, n_{gs} = n_{bd}\,,\text{and}\,,
n_{ru} = n_{gd} = n_{bs}\,.
$
A sum over color indices for up quarks gives:
$
n_u = n_{ru}+n_{gu}+n_{b u} = n_{ru} + n_{rd} + n_{rs} = n_r\,,
$
where we have used the identities from above.
Analogously, we find~$n_d = n_g$ and~$n_s = n_b$.
Color neutrality requires~$n_r = n_g = n_b$, which together implies~$n_u = n_d = n_s$. Given the electric charges of the quarks, this ensures that their contributions cancel exactly, making the system electrically neutral.
Therefore, no electrons are needed, and the electron chemical potential vanishes:~$\bar\mu_{\rm e} = 0$.

\section{Finite strange-quark mass}
\label{app:strangequarkmass}
Here, we detail the evaluation of the strange-quark mass correction~$\sim \bar m_s^2 \barDeltaCFL^2$ to the pressure.
We start from the quark contribution to the effective action in the color-superconducting phase for finite quark masses at one-loop order. 
The corresponding expression can be deduced from Eq.~(5.18) in Ref.~\cite{Buballa:2003qv} by restricting ourselves to the CFL condensate and setting the masses of the up and down quarks to zero. 
To order~$\bar m_{\rm s}^2$, we then find that the strange-quark mass correction to the effective action to LO in the diquark field~$\Delta$ is given by 
\begin{align}
\frac{\Gamma_{\ms^2}^{1\rm-loop}}{V_4} = - \ms^2 \Delta^{2} \frac{6 \ln \Delta^{2} + 9 - 8 \ln 2}{6 \pi ^2} + \mathcal{O}(\Delta^3)\,,
\end{align}
where~$V_4$ is the spacetime volume.
The minimization of the effective action with respect to the diquark field as performed in~\cite{Geissel:2024nmx} eventually yields the term~${\sim \bar m_{\rm s}^2 \barDeltaCFL^2}$ in our expression for the pressure in Eq.~\eqref{eq:pressureNLO}.

\section{Feynman rules\label{app:FeynmanRules}}
For the computation of the effective action in the presence of a CFL gap in the quark propagators, it is convenient to define Feynman rules as usually done in perturbative computations in QCD.
Since terms of the form~$\sim \alpha_{s} \barms^2\barDeltaCFL^2$ are beyond the order considered in the present work, as they are a product of three small quantities, it suffices to restrict ourselves to the chiral limit here; i.e., we set all quark masses to zero.
In addition, we can also set all quark chemical potentials to be equal. 
The calculation of corrections associated with a nonzero strange-quark mass at~$\mathcal{O}(\alpha_{s}^0 \bar m_{\rm s}^2\bar\Delta_{\rm CFL}^2)$ is discussed in App.~\ref{app:strangequarkmass}.

In the chiral limit and for equal quark chemical potentials, the quark propagator matrix assumes the form

\begin{align}
\mathcal{P}_\psi & \equiv
\begin{pmatrix}
\langle \psi^T(-P) \psi(Q) \rangle        & \langle \bar\psi(P) \psi(Q) \rangle        \\
\langle \psi^T(-P) \bar\psi^T(-Q) \rangle & \langle \bar\psi(P) \bar\psi^T(-Q) \rangle \\
\end{pmatrix}
\nonumber
\\
& =
\begin{pmatrix}
V_\psi & X_\psi \\
Y_\psi & W_\psi \\
\end{pmatrix}
\deltapq{P-Q} \,,
\label{eq:propmatrix}
\end{align}
where the off-diagonal entries of the quark propagator matrix are given by
\begin{align}
X_\psi = (Y_\psi^*)^T & = - \left(\slashed{P}_+ +  \Delta^{2} \frac{\slashed{P}_-}{P_-^2}\right)G_{\psi\Delta}^+ \delta^{ij}\delta_{ab}
\nonumber\\
& \hspace{0.4cm} + \Bigg[\left(\slashed{P}_+ +  \Delta^{2} \frac{\slashed{P}_-}{P_-^2}\right)G_{\psi\Delta}^+
\nonumber\\
& \hspace{0.75cm} - \left(\slashed{P}_+ + 4\Delta^{2} \frac{\slashed{P}_-}{P_-^2}\right)G_{\psi\Delta}^{{\rm CFL},+}
\Bigg]
\frac{\delta_{a}^i\delta_{b}^j}{3} \,.
\end{align}
Here,~$i,j$ are flavor indices whereas~$a,b$ are color indices.
For convenience, we introduced the momentum~${P_\pm \equiv (P_0 \pm {\rm i} \mu, \vec{P}^{\,})^T}$, where~$\mu = \mu_{\rm B}/3$ is the quark chemical potential. 
Note that the quark propagators depend on the diquark field $\Delta$. 

Due to the presence of a gap in the excitation spectrum of the quarks, the quark propagator matrix also comes with nonzero diagonal elements, 
\begin{align}
V_\psi = - W_\psi^* &= \Delta\Bigg\{ \left(P_+ \cdot P_- +  \Delta^{2}\right)G_{\psi\Delta}^+ G_\psi^- \epsilon_{abA} \epsilon^{ij A}
\nonumber\\
&\hspace{0cm}- \bigg[\left(P_+ \cdot P_- + \Delta^{2}\right)G_{\psi\Delta}^+
\nonumber\\
&\hspace{0cm}- \left(P_+ \cdot P_- + 4\Delta^{2}\right)G_{\psi\Delta}^{{\rm CFL},+}
\bigg]G_\psi^-
\frac{2}{3}\delta^i_a \delta^j_b \Bigg\} \gamma_5 \mathcal{C}\,.
\end{align}
For the sake of readability, we have introduced the following quantities:
\begin{align}
G_\psi^\pm          & \equiv \frac{1}{P_\pm^2},
\\
G_{\psi,\Delta}^\pm & \equiv \frac{P_\mp^2}{P_\pm^2P_\mp^2 + 2 \Delta^{2} P_\pm \cdot P_\mp+ \Delta^{\!4}}\,,
\\
G_{\psi,\Delta}^{{\rm CFL},\pm} & \equiv \frac{P_\mp^2}{P_\pm^2P_\mp^2 + 8 \Delta^{2} P_\pm \cdot P_\mp+ 16 \Delta^{\!4}}\,.
\end{align}
Since the gap entering the gluon propagator leads to contributions to the effective action that are of higher order than considered in the present work~\cite{Geissel:2024nmx}, we use the bare gluon propagator in Feynman gauge
\begin{align}
\left(\mathcal{P}_A^{0}\right)_{\mu\nu}^{ab}
=
\frac{1}{P^2} \delta^{ab}
\delta_{\mu \nu} \deltapq{P-Q} \,,
\end{align}
in our computations.
Finally, the quark-gluon vertex is parametrized as
\begin{align}
(\Gamma^{(3)})_{bc,\mu}^a  =
\begin{pmatrix}
0                           & - \bar{g} \left(T_{bc}^a\right)^T \gamma_\mu^T \\
\bar{g} T_{bc}^a \gamma_\mu & 0                                              \\
\end{pmatrix}\,.
\end{align}

\section{Two-loop contribution}
\label{app:twoloopcontribution}
Within this expansion scheme, the integral expression of the two-loop diagrams in Eq.~\eqref{eq:2loopdiagamsapprox} is given by
\begin{widetext}
	\begin{align}
	\frac{\Gamma^{2-\text{loop}}_\text{quark}}{V_4}
	& = \frac{1}{2} \int_{P,Q} \left[\mathcal{P}_A^0\right]^{a a'}_{\mu\nu} \left(P-Q\right)
	\Tr \Big\{ (\Gamma^{(3)})_{bb',\nu}^{a'} \left[\mathcal{P}_\psi^\Delta\right]_{b' c}^{ij}\left(P\right) (\Gamma^{(3)})_{cc',\mu}^{a}
	\left[\mathcal{P}_\psi^0\right]_{c' b}^{ji}\left(Q\right) \Big\} \,.
	\end{align}
\end{widetext}
These terms are explicitly of~${\mathcal O}(\Delta^{2})$.
As a consequence of our approximation, contributions originating from the diagonal propagator elements~$V_\psi$ and~$W_\psi$ in Eq.~\eqref{eq:propmatrix} vanish up to this order since~$V_\psi|_{\Delta = 0}  = 0 = W_\psi|_{\Delta = 0}$.
However, we note that terms of the form~$\mathcal P_A^0 \Gamma^{(3)} V_\psi \Gamma^{(3)} W_\psi$ are also of~${\mathcal O}(\Delta^{2})$.
Evaluating the traces, we can bring the above integral into a similar form as in Ref.~\cite{Geissel:2024nmx}.
For the two-loop correction, we eventually find
\begin{align}
\!\!\!\!\frac{\Gamma^{2-\text{loop}}_\text{quark}}{V_4}
= 128\pi \alpha_s \Delta^{2} \mu^2
\left[-2.44 + 0.0078 \ln\left(\frac{\Delta^{2}}{\mu^2}\right)\right]
\label{eq:gamma2loop}\,.
\end{align}
\end{document}